# Analysis of Software Delivery Process Shortcomings and Architectural Pitfalls

Amol S Patwardhan

Abstract: This paper highlights the common pitfalls of overcomplicating the software architecture, development and delivery process by examining two enterprise level web application products built using Microsoft.Net framework. The aim of this paper is to identify, discuss and analyze architectural, development and deployment issues and learn lessons using real world examples from the chosen software products as case studies.

Introduction: Software industry is undergoing massive globalization. Outsourcing and offshoring is not limited to the traditional information technology service industries in Asia. In fact, many software companies are hiring contractors from Mexico, Chile, East Europe, Vietnam and Bangladesh. Typically, the software development retains the core competencies (architects, team leads) and resources with business and technical knowledge and hand over the software maintenance tasks (bug fix, small enhancements, production environment troubleshooting, minor projects) to the offshore teams. These teams consist of developers with varying English and programming language skillset. With the increasing diversity of teams working on the code base it is critical to have a robust, consistent and easy to maintain architecture and software framework. (G. Abowd, 1997) has examined various software architecture evaluation practices for the software industry.  (M. A. Babar & I Gorton, 2004) have studied scenario based software architecture evaluation methods. (M. Barbacci et. al, 1998) have done trade off analysis on software quality attributes. (P. Bengtsson & J. Bosch) have studied prediction of maintainability of software architecture. (P. Clement, R. Kazman & M.Klein, 2002) have provided methods and case studies for evaluating software architecture. (C. H. Lung & K. Kalaichelvan, 2000), (N. Lassing, D. Rijsenbrij & H. V. Vliet, 1999), (D. L. Parnas & D. M. Weiss, 1985) have examined architecture sensitivity, reviewed design practices and performed risk assessments on software development, delivery and architecture. (M. H. Klein et. al, 1993) and (M. R. Lyu) have analyzed real time software reliability and made recommendations in their handbooks for process improvements. Most of these studies have performed the analysis in the late nineties or early part of this century. Since then software industry has made technology advancements in terms of web development, client-server architecture, infrastructure availability (network speed, disk space and processor speeds). The nature of software development in terms of team size and location has also changed drastically. As a result, this paper tries to evaluate two enterprise level web applications built in recent years with a more diverse development team structure.

Method: Two enterprise level web application products were chosen for comparative analysis. A series of questions was constructed and the software development and deployment process maturity and architectural quality was evaluated. Pitfalls and challenges in maintaining software consistency, architectural conformance and software delivery processes were identified. The author served as a senior level .NET consultant on both the products and was familiar with the development process, code and other information technology and operations departments in both organizations.

Case Study 1: The first product (code name Product A, company name A) was a web application built using C# ASP.NET web forms. The web application was based on n-tier architecture. It used windows services and message queues (MSMQ) at the backend for asynchronous batch processing. The backend database used MS SQL 2008. The web application also interacted with external vendor systems using a legacy program based on C++ and WCF based web services.

Case Study 2: The second product (code name Product B, company name B) was a web application built using C# ASP.NET model view controller (MVC) framework. The web application was based on micro-services architecture and interacted with several internal legacy systems built using MS Access and VB. The backend database used MS SQL 2008. The system also interacted with a JAVA based system and external vendors through WCF web services, SQL batch jobs and scheduled jobs.

The next section contains the questionnaire used for evaluation of the software product, development and deployment process:

1) How many environments is the code deployed?

Product A: Was deployed in development environment. If automated smoke tests passed the code was deployed to quality assurance department (QA) or the test environment. Once the feature was signed off the code was deployed to the user acceptance testing (UAT) environment. Once the feature was signed off by the business analyst, the code was pushed to the production environment.

Product B: Was deployed in the development environment. If the sanity testing and smoke testing was signed off, it was pushed to the QA environment. Once the testing was signed off, UAT was performed on the same environment. Finally, the code was deployed to the staging (silo) environment and once it was signed off in the silo environment it was pushed to production. Product B also supported concurrent releases adding to the complexity.

In both cases there were too many environments to maintain and the waterfall software development methodology meant, longer release duration for features and bug fix. Every time the code was deployed to a certain environment, significant time and resources had to be dedicated for troubleshooting environment issues.

2) How many load balanced servers exist in each tier?

Product A: Three load balanced servers were used. The sessions were sticky. A request coming to server 1 went to the application server 1 and then to the database. Disadvantage: This meant the load balancing was not uniform leading to less than ideal performance.

Product B: Three load balanced servers were used. The sessions were not sticky, which meant a request coming to a webserver could flow to any of the three load balanced application servers. Disadvantage: Troubleshooting a production issue was trickier since there was no good way to quickly track the error.

3) How many tiers exist in the deployed software?

Product A: Consisted of web tier (3 servers), application server (3 servers), service layer (3 servers) and databases tier (3 servers).

Product B: Consisted of web tier (3 servers), application server (3 servers), service layer (3 servers), database tier (3 servers).

4) Is the code deployed on a shared fileserver or on individual nodes?
Product A code was deployed on a file share server. Multiple webservers on different tiers pointed to the same code base. Advantage: The code deployment needed to be done only in one location. This resulted in maintenance of fewer configuration file and lesser deployment errors. Disadvantage: All the servers pointed to the same location, causing contention issues, file locking and also the risk of single point of failure. Although there was backup and server restoration mechanism in place, but even then there was a risk of down time while the system came back up to full functionality.
Product B code was deployed locally on each tier. As a result, paths, flags and other configuration changes were prone to errors such as human errors in manual deployment and errors due to access/ permission issue during automatic deployment.

5) Is the code easy to debug?
Product A code contained proprietary and custom frameworks built on decade old .Net technology making the debugging process cumbersome.
Product B code contained several micro services making the debugging process lengthy. The programmers had to setup break points in multiple instance of the service specific code files and projects.

6) Is the software web application easy to setup on a local developer machine?
Product A code did not have automated scripts to assist in the local setup.
Product B code used PowerShell scripts to automate most of the local setup process.

7) Is the deployment automated?
Product A used a custom deployment tool with very little documentation.
Product B used a custom deployment tool with very limited documentation. Not all the processes had been automated and thus the deployment was prone to human error.

8) How long does the deployment take?
Product A build and deployment tool 2-3 hours. The build and deployment was lengthy and the automated smoke tests, database backup and restore process and application of database build scripts, took too long.
Product B build and deployment took less than 30 min. There was no backup, restore and reapplication of database build script.

9) Does the architecture contain redundant libraries and framework?
Product A contained business layer and a newly written custom xml based framework. But the old business layer was not completely removed.
Product B contained several obsolete third party libraries and configuration files that could be potentially cleaned up.

10) Does the architecture contain redundant tiers?

Product A contained externally available web services on a separate tier which could be combined with the web application tier. There was also a separate service layer that communicated with the web services which could be combined with the application tier.
Product B contained an externally available service layer on its own tier which could be combined with the web tier.

11) Does the architecture use new technologies that were embraced to stay current with the market hype?
Product A used BizTalk for business to business communication which led to high licensing call and maintenance effort. The same communication could have been implemented using custom web services.

12) Does the architecture contain inconsistencies in naming variables?
Both Product A and B suffered from several inconsistencies in naming variable in the code which depended on the programmer preference instead of enforced coding standard.

13) Does the architecture contain inconsistencies in naming functions?
Both Product A and B suffered from several inconsistencies in naming functions in the code which depended on the programmer preference instead of enforced coding standard.

14) Does the architecture contain inconsistencies in design patterns?
Both Product A and Product B suffered from architectural erosion and lack of consistency in layout of classes, inheritance structure, project and solution organization folders were inconsistent in terms of location of code for user interface, business code, utility, helper classes and data layer classes.

15) Does the architecture contain unit tests coverage above 70%?
Neither Product A or Product B contained significant number of unit tests and the code coverage was minimal. This showed a general aversion towards writing unit tests. The developers either had no time to spend on meaningful unit tests or invest in maintaining unit tests that were no longer valid because of changes to the code.

16) Does the architecture contain legacy code?
Product A was all built on ASP.net web forms technology. So it was difficult to fully transition to MVC based web application.
Product B contained a lot of legacy code ranging from asp pages, web forms and an intermix of C# and VB code.

17) Does the development shop allow local instances of the database?
Product A allowed developer to work on local instances of the database containing dummy and test data.
Product B was developed using a common development environment.

18) Is the production data available for troubleshooting?

Product A had provision to restore scrubbed copy of production database upon request.
Product B had an environment that contained restored copy of previous day's production data.

19) Does the troubleshooting team have access to production environment?
Troubleshooting team for Product A did not have direct access to run queries or look at error logs in the production environment. As a result, the team had to rely on technical resources with sufficient access to provide the necessary data. This led to longer response times to production issue.
Troubleshooting team for Product B had limited access to production servers and database and error logs. This led to better service response times but at the risk of someone accidentally changing configuration in the production environment or running a long running query.

20) Does the support team have access to tools (app dynamics, splunk) for easier troubleshooting?
Both Product A and B support teams used popular industry tools to troubleshoot production and environment issues.

21) Is there detailed documentation about various modules in the software?
Both Product A and B lacked detailed technical documentation about code and various modules in the software.

22) Is there a coding standards document available?
Product A team did not have a coding standard document while Product B team had an official coding and database standard document.

23) Is the code review process followed and enforced?
Product A development team did not have a coding standards document and the architects used their expertise to enforce coding standards during code reviews. This resulted in less than ideal code review quality because occasionally reviews would be delegated to senior developers who were busy and did not check the code diligently.
Product B development team had a coding standard document but lacked proper process to enforce the standards on various geographically diverse teams who relied on their team leads to enforce the standards.

24) Do the support team get timely help from development team during troubleshooting effort?
Product A support team had to rely to raise a ticket and wait until the ticket was assigned to a development resource to help with a troubleshooting effort.
Product B support team had active involvement from development team resources to aid in troubleshooting process.

25) Does the support team get guidance from the development team while working on a feature?
Both Product A and Product B support teams received high level guidance from subject matter experts on the development teams.

26) Does the inter-departmental communication suffer from longer response times?
Organization A and B had several IT operation departments with mismatching priorities. As a result, it was difficult and time consuming process to get immediate help from various departments.

27) Does the development team have a support ticket system?
Product A and B development teams used support ticket system such as Microsoft CRM and Dell Kace.

28) Does the development team use a code repository?
Both Product A and B development teams used TFS 2010 as the code repository management system.

29) Does the development team use a main and project branch as development strategy?
Both product A and B development teams had a main branch to deploy code to the production environment and project branches for feature development.

30) Does the architecture contain modules integrating with 3rd party vendor systems?
Both Product A and B contained custom code specifically written for integration with third party system. But both the products lacked sufficient documentation and only a few subject matter experts on these modules.

31) Does the development team receive timely response from the 3rd party vendors?
Product A and Product B did not have a point of contact and development resource dedicated to communicate with 3$^{rd}$ party vendors. This caused confusion among developers over who would communicate and negotiate and obtain the application programming interface (API) documents from the vendors during feature additions. This resulted in delays in communication.

32) Is the architecture upgraded to the latest version of the framework?
Product A was on version 4.0 of .NET and Product B was on version 4.5.

33) Does the architecture contain legacy, 3rd party assemblies preventing the project from upgrading to a newer version?
Product A and Product B had some modules which could not be upgraded because of dependency on 3$^{rd}$ party libraries which did not support latest version of framework.

34) Does the architecture contain inconsistent client side libraries?
Both Product A and Product B contained mixed versions of client side library versions of jQuery, jQuery UI, jQuery mobile. It was important to upgrade all the libraries to the latest version to avoid inconsistent client side behavior.

35) Does the architecture differ across different products in the same organization?

Product A and Product B contained different design patterns within their own code base and some of the backend tools did not follow any specific design pattern.

36) Does the code base contain dead code?
Both Product A and Product B contained several lines of commented code that could be cleaned up.

37) Does the code contain sufficient comments in critical and complex modules of the software?
Both Product A and Product B contained several complex functions that did not have sufficient comments or documentation about the logic and the resources who had worked on the code or subject matter experts had either left the organization or moved to other project or department.

38) Do the development team have a dedicated database team or are developers responsible for writing their own SQL?
Product A did not have a dedicated database team and relied on the developers writing their own SQL. Product B had a dedicated database team to enforce database standards.

39) Is the code analyzed using code analysis tool?
Product A and Product B were analyzed for code complexity, common industry standards and .Net coding standards.

40) Is the code inspected for security holes?
Both Product A and Product B were evaluated by third party software for commonly known software industry specific security vulnerabilities.

41) Is the code obfuscated?
Neither Product A or B obfuscated the code.

42) Is the code deployed onsite or client site?
Product A supported client-server model as well as client site only installations. This led to higher installation cost in terms of engaging resources to support client installs.
Product B only supported client-server installation model.

43) Do all the projects have requirement, design and test scenario documents?
Product A and B used waterfall software development methodology. Product A was also audited every year for compliance to industry standard and hence had requirements, design and test models well documented. Product B did not have sufficient documents for all projects in development.

44) Does the software use load balancing hardware?
Product A and B used load balancing hardware.

45) Does the software handle authentication use custom authentication or hardware?
Product A used custom forms based authentication and the users were authenticated against the database.
Product B used external hardware to handle authentication and redirection of the web request to the web site. Disadvantage: This led to dependency on the infrastructure team to implement feature enhancements in the login process.

46) Does the code contain sufficient error logging?
Product A used custom error logging in the database.
Product B used custom windows error logging.

47) Does the code contain sufficient checkpoint logging?
Product A contained insufficient logging to track program execution.
Product B contained relatively better coverage in terms of function entry and exit points.

48) Does the code contain single solution or multiple solutions?
Product A contained 3 solutions. Product B contained several solutions because it was based on micro services.

49) Does the development team have a performance and load testing environment?
Product A had a performance and load testing environment but it was not well maintained and had poor hardware.
Product B did not have an automated load and performance testing environment.

50) Does the development team invest in onboarding new hires?
Product A had poor onboarding process with a clear disconnect between hiring managers and development team in terms of introduction, new hire training and task assignment in the first few weeks.
Product B had a well formulated on boarding process ranging from new hire training and dedicated mentors during the initial few weeks for the employee.

51) How many architects/ senior software engineers from the original team are still with the organization?
Company A could not retain a single architect from the original team who had worked on the inception of the Product A.
Company B had two architects who were still with the development team right from the inception phase of Product B.

52) How many times has the company changed ownerships?
Company A and B had changed ownership twice in the last decade.

53) How many times has the work culture got affected by change in owners?
Work culture in organization for Product A and B was affected because of change of ownership twice in the last decade.

54) Have there been layoffs and has it adversely affected the IT department?
Organization for Product A had suffered layoffs thrice in the last decade due to low revenue. Organization for Product B had layoffs once in the last decade due to low revenue.

Conclusion: Development of enterprise level software and maintaining it with help from offshore teams could be made more cost effective if the software has sufficient documentation for critical modules, training material, code comments and consistent architecture. The research found too many flavors of architecture and design patterns, lack of enforcement of coding standards and poor code coverage using unit testing. The reasons given by development resources were looming deadlines, lack of budget for internal IT initiatives to pay off technical debt and knowledge transfer and architectural improvements. The research paper attempted to highlight the development, deployment and architectural pitfalls so that future software products can learn what to do and what not to do from this article. The paper also presented a series of questions that can be asked to assess the overall health of a software product as opposed to quantitative evaluation using metrics like maintainability, testability, unit tests, code complexity and comments used by some of the state of the art techniques. This paper focused on two real world enterprise level examples from software industry in an attempt to identify the challenges in typical software development shops that need to be supported by geographically diverse teams.